\documentclass[prl,twocolumn]{revtex4}\usepackage{psfrag}
\usepackage{epsfig}
\usepackage{amsmath}
\usepackage{amssymb}

\usepackage[colorlinks=true,linkcolor=blue,citecolor=blue]{hyperref}

\newcommand{\bra}[1]{\ensuremath{\langle#1|}}
\newcommand{\ket}[1]{\ensuremath{|#1\rangle}}

\newcommand{\be}{\begin{equation}}
\newcommand{\ee}{\end{equation}}
\newcommand{\avg}[1]{\ensuremath{\langle #1 \rangle}}
\newcommand{\aver}[1]{\ensuremath{\langle #1 \rangle}}

\newcommand{\ie}{{\it i.e.}}

\newcommand{\eg}{{\it e.g. }}

\newcommand{\etal}{{\it et al.}}

\newcommand{\JEFF}{\hat{\mathcal{J}}}
\newcommand{\SOP}{\hat{\mathbf{j}}}

\begin{document}

\title{\bf Quantum Memory Assisted Probing of Dynamical Spin Correlations}

\author{O. Romero-Isart$^{1}$}
\author{M. Rizzi$^{1}$}
\author{C. A. Muschik$^{1}$}
\author{E. S. Polzik$^{2}$}
\author{M. Lewenstein$^{3,4}$}
\author{A. Sanpera$^{3,5}$}

\affiliation{$^1$Max-Planck-Institut f\"ur Quantenoptik,
Hans-Kopfermann-Strasse 1,
D-85748, Garching, Germany.}
\affiliation{$^2$Niels Bohr Institute, Danish Quantum Optics Center
 QUANTOP, Copenhagen University, Blegdamsvej 17, 2100
Copenhagen Denmark. }
\affiliation{$^3$ICREA--Instituci\'{o} Catalana de Recerca i Estudis
Avan\c{c}ats, E-08011 Barcelona, Spain. }
\affiliation{$^4$ICFO-Institut de Ci\`{e}ncies Fot\`{o}niques,
Mediterranean Technology Park, E-08860 Castelldefels (Barcelona),
Spain. }
\affiliation{$^5$ Departament de F\'{i}sica.  Universitat Aut\`{o}noma de Barcelona, E-08193
Bellaterra, Spain.}

\begin{abstract}

We propose a method to probe time dependent correlations of non trivial observables in many-body ultracold lattice gases.
The scheme uses a quantum non-demolition matter-light interface, first, to map the observable 
of interest on the many body system into the light and, then, to store coherently such information into an external system acting as a quantum memory. 
Correlations of the observable at two (or more) instances of time are retrieved 
with a single final measurement that includes the readout of the quantum memory.  Such method brings at reach the study of 
dynamics of many-body systems in and out of equilibrium by means of quantum memories in the field of quantum simulators.

\end{abstract}

\maketitle

The quantum simulation of many-body physics with ultracold
atoms in optical lattices requires the ability to prepare,
manipulate, and probe the states~\cite{ReviewsColdAtoms}. 
Seminal experiments in this field range from the realization of a Bose-Einstein condensate
with alkali atoms in the weak-coupling regime~\cite{Anderson1995,Davis1995},
to the Mott-insulator--superfluid transition in the strongly interacting 
regime~\cite{Jaksch1998,Greiner2002}.
Due to the high degree of control offered by ultracold lattice gases, simulation of quantum magnetism 
is becoming one of the most ambitious goals in the field~\cite{Trotzky2008, Magnetism,Simon2011}. 
Candidate systems are made up by optical lattices where the particles (either bosons or fermions)
arrange themselves in a deep Mott insulator, 
and the internal degrees of freedom of the atoms play the role of spins.
Such internal states might be given from hyperfine structure in the case of alkalis~\cite{ReviewsColdAtoms}
or by nuclear spins for alkaline-earth fermions \cite{Hermele2009}.
At sufficiently low temperatures and entropies,
the spin-spin interactions arising perturbatively from super-exchange processes (virtual tunneling between neighboring sites)
are predicted to give rise to quantum magnetism phenomena as Ne\'el ordering, SU(N) magnetism, spin Hall effects, 
and Stoner magnetism, see \eg \cite{Sachdev2008}.

Quantum simulation with cold gases does not only aim at
mimicking the phenomena encountered in condensed matter physics, but
also at exploring new frontiers in physics. 
In particular, ultracold atoms permits to explore non-equilibrium phenomena in 
{\it closed} systems ~\cite{Cazalilla2010}.
Within this context, the study of thermalization in out-of-equilibrium situations after a quench, 
the influence of metastable states during the dynamics, the effect of quantum correlations out of equilibrium, 
or the role the steady states are few of the open questions. To address these questions, efficient probling tools 
are necessary ~\cite{BraggSpectroscopy}.  


Here we propose to probe dynamical correlations in strongly interacting ultracold gases by combining   
a quantum non-demolition matter-light interface with a quantum memory. 
The idea, stemming from quantum repeaters, exploits the well known fact  
that light is a good carrier of information however difficult to store, while matter can be made to store coherently information for long times, serving
thus as an efficient quantum memory.  Specially suited for such a goal is the QND Faraday interface which maps 
the spin-polarization of atoms into the polarization of photons.  Such QND measurement has been recently implemented to spin squeeze 
states in cold atomic ensembles well below the projection noise limit ~\cite{Appel}. It has also been presented as an efficient tool to probe quantum magnetism in ultracold atoms as proposed in~\cite{Eckert2007,Eckert2007b,Mekhov2009,Roscilde2009a,DeChiara2010}  and experimentally demonstrated 
~\cite{Esslinger2011}. In our proposal, the QND interface is used both, to coherently map a relevant operator of the many-body system $\JEFF$ at different times ($t$, $t+\tau$,~\ldots) to the light  {\em without} performing a measurement at each time, and also to store $\JEFF$ into a quantum memory  between two consecutive matter-light interactions for a time $\tau$ comparable to the time scale of the internal many-body dynamics.
A final single measurement which includes a readout of the memory yields dynamical spin correlations $\bra{\psi}\JEFF(t) \JEFF(t+\tau) \JEFF(t+ \tau') \ldots \ket{\psi}$, for some given state $\psi$. This quantum memory-assisted probing (QMAP) scheme is minimally destructive, does not depend on the linear response of the system to an external perturbation~\cite{Chaikin}, and is equally suitable for systems out of equilibrium, \ie~when $\psi$ is not an eigenstate of the many-body Hamiltonian. We show that the correlations obtained using QMAP are fundamentally different from those obtained by repeating a QND measurement of $\JEFF(t)$ at different instances of time and then performing a correlational analysis. This difference comes from the quantum interference effect present in the QMAP process, since a single measurement is performed after coherently storing the observable of interest at different times.

Before explaining in detail our protocol, let us first review the basic elements of a Faraday QND interaction. 
When a strongly polarized light beam interacts off resonantly with the internal spin degrees of freedom of an atomic system  the polarization of the photons rotate, an effect known as the Faraday rotation (see the recent review~\cite{Hammerer2010} and reference therein).  If the light beam is e.g. strongly polarized along the $x$-axis and propagates along the $z$-axis, it can be fully described by time-integrated canonical operators $\hat X_L=\hat S_2/\sqrt{N_\text{ph}}$ and $\hat P_L=\hat S_3/\sqrt{N_\text{ph}}$, where $\hat S_2$ ($\hat S_3$) is the Stokes operator corresponding to the difference in the number of photons in the $\pm 45$ (in the two circular) polarizations, and $N_\text{ph}$ is the total number of photons in the beam. After the Faraday interaction, the integrated equations of motion result into~\cite{Hammerer2010,Eckert2007b}
\be \label{eq:F}
\hat X^{\text{out}}_L =\hat  X^{\text{in}}_L - \kappa\JEFF(0),
\ee 
where $\hat X^{\text{in}}_L$ ($\hat X^{\text{out}}_L$) is the light quadrature before (after) the interaction 
The coupling strength can be expressed as $\kappa=\sqrt{d \eta_{A}}$, where $d$ is the optical depth of the atomic sample, and $\eta_{A}$ is the spontaneous emission probability induced by probing.
An optimal QND interaction is achieved
with $\eta_{A}$ in the range from $ \approx d^{-1/2 }$ for single color probing to the constant $ \approx 1/3$ for two-color probing~\cite{Hammerer2010}. For cold samples $d\ge 100$~\cite{Schnorrberger2009} and hence $\kappa^2 \approx 10$ should be within reach.   The observable
\be \label{eq:J}
\JEFF= \frac{1}{\sqrt{N}} \sum_n c_n \hat j_n^z,
\ee
 corresponds to the total modulated magnetization of the atoms illuminated by the light beam. $N$ denotes the total number of atoms confined in an optical lattice, and $\hat j_n^z$ the $z$-component of the atomic spin at site $n$.
 The modulation, given by the coefficients $c_n$, reflects the spatial dependence of the light beam intensity; in a standing wave configuration~\cite{Eckert2007b},  $c_n= 2 \cos\left(k n a-\alpha \right)^2$, where $k$ is the wave number of the probing laser, $a$ is the optical lattice spacing, and $\alpha$ describes the shift between the probing standing wave and the optical lattice.

The many body operator $\JEFF$ is a QND observable~\cite{Braginsky1992} since it commutes with the Faraday Hamiltonian used for the measurement. 
As required to measure non-trivial dynamical correlations, $\JEFF$ does not commute with the many-body Hamiltonian. The light-matter interaction time can be chosen in the $\mu$sec range~\cite{Appel}, \ie, much shorter than the relevant timescale of the many-body system and be considered instantaneous, so that the QND character of the interaction is preserved. As shown in~\cite{DeChiara2010}, $\JEFF$ corresponds to a non-trivial observable of magnetic systems. The aim of this work is to use the QND probing scheme to access to correlations functions as $\bra{\psi} \JEFF(t) \JEFF(0) \ket{\psi}$ for any initial state $\psi$.
We concentrate on two-time correlations to explain the method but extensions to $n$-time correlations are straightforward.

\begin{figure}[pbt]
\begin{center}
\includegraphics[width=\linewidth]{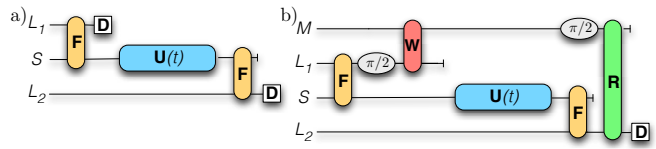}
\caption{(Color online) Schematic representation as a quantum circuit of the protocol to measure dynamical correlations. S (M) is the many body system (the quantum memory), and $L_{i}$ is the $i$-th light beam used in the protocol. The definition of the quantum gates is provided in the text. a) Scheme for measuring $\JEFF$ at two different instances of times, which after repetition give $F_{S}(t)$. b) Scheme to measure $F_{M}(t)$  by using a quantum memory.}
\label{Fig:Scheme}
\end{center}
\end{figure}

To better understand the role of the quantum memory, let us first analyze the information acquired by just performing 
independent QND measurements of $\JEFF$ at different times (without storing them in a memory) and making a statistical analysis of the results.
Suppose our many-body system is initially prepared in 
some state $\ket{\psi}$. At $t=0$, a measurement of $\JEFF=\sum_i a_i \hat P_i$ is performed, where $\hat P_i$ projects into the eigenspace with eigenvalue equal to $a_i$. Depending on the outcome $a_i$, the state collapses to $\hat P_i \vert \psi \rangle$.
In our set-up, this measurement corresponds to performing a QND Faraday interaction between the system and the light, 
followed by a homodyne measurement of the relevant light quadrature.
After the measurement, the many-body system evolves $\hat U(t)=\exp[- i \hat H_\text{MB} t]$ under the many body Hamiltonian, $\hat H_\text{MB}$, during a time window
of length $t$ until a new measurement of $\JEFF$ is performed yielding the outcome $a_j$, as schematically sketched in Fig. 1a. The most general expression for the statistical mean resulting from both outcomes is given by
\be
\begin{split} \label{eq:signalNOK}
F_S(t)= \sum_{i} a_i  \bra{\psi}  \hat P_i \JEFF(t) \hat P_i \ket{\psi}.
\end{split}
\ee
This quantity depends explicitly through expression (\ref{eq:J}) on the parameters $k$ and $\alpha$ of the probing laser. If the state of interest is prepared in the ground state of the many-body system,  {\em i.e.} $\ket{\psi}=\ket{E_0}$, the Fourier transform of $F_S$  reads $C_{S}(\omega)= \int dt e^{\text{i} \omega t} F_{S}(t) = \sum_{i,j} \xi_{ij} \delta(\omega - (E_i-E_j))$, where $\{E_{i}\}$ is the energy spectrum and the amplitudes $\xi_{ij}$ can be trivially obtained. This quantity signals the frequencies corresponding to energy differences between different eigenstates, providing {\it partial} information 
about the energy spectrum.
Although related, $F_{S}(t)$ is not the two-time correlator of the observable $\bra{\psi}\JEFF(t) \JEFF(0)\ket{\psi}$.

To obtain two-time correlation functions we use optical quantum memories~\cite{Lvovsky2009} based on atomic ensembles~\cite{Hammerer2010}.
In the atomic memories, atoms, analogously to light, are strongly polarized along one direction, such that the spatially integrated spin components in the orthogonal plane, $\hat X_M(t)$ and $\hat P_M(t)$,  fulfill canonical commutation rules. The Faraday interaction between the light and the quantum memory yields the relation
\be
\hat X^\text{out}_M=\hat X^\text{in}_M + \kappa_W \hat P^\text{in}_L,
\ee
where $\hat X^\text{in}_M=\hat X_M(t=0)$ and $\hat X^\text{out}_M$ is the output value after the interaction with the light beam. The light quadrature $\hat P^\text{in}_L$ is $\hat P_L(z=0)$. Note that both light and matter quadratures can be rotated according to $\hat X \rightarrow \tilde X=\cos(\phi) \hat X + \sin(\phi )\hat P  $ and  $\hat P \rightarrow \tilde P=\cos(\phi) \hat P- \sin(\phi )\hat X $.
The quadrature stored in the memory can be retrieved using a second light beam and the corresponding Faraday interaction, such that
$\hat X^\text{out}_L=\hat X^\text{in}_L + \kappa_R \hat P^\text{in}_M$ . Such type of memories have been experimentally realized with atomic samples at room temperature in~\cite{Julsgaard2004}, where a storage time of the order of $4$ ms was reported. Alternatively, memories with ultracold atoms in optical lattices have achieved longer storage times up to $240$ ms~\cite{Schnorrberger2009}. These time storages are typically large enough to address the evolution of many-body ultracold gases. 
 
The protocol leading to spin dynamical correlations using a Faraday probing and a quantum memory ---sketched in Fig.~\ref{Fig:Scheme}b---, 
is straightfoward and requires some rotations of the quadratures to obtain finally the correct time correlation. It is summarized as follows:\\
\noindent (i) At $t=0$, a Faraday interaction (F) between the many body system S and the first light pulse L$_1$ maps the (modulated) magnetization of the
system $\JEFF$ to the light quadrature according to 
$\hat X^{\text{out}}_{L_1}  =  \hat X^\text{in}_{L_1} - \kappa_1\JEFF (0)$ and
$\hat P^{\text{out}}_{L_1}  =  \hat P^\text{in}_{L_1}$. A rotation of $\phi=\pi/2$ is applied to the output light quadrature.\\
\noindent (ii) The rotated quadrature of this first light beam is mapped into the quantum memory M according to:
$\hat X^\text{out}_M = \hat X^\text{in}_M + \kappa_W \tilde P^{\text{out}}_{L_1}
=\hat X^\text{in}_M - \kappa_W ( \hat X^\text{in}_{L_1} - \kappa_1\JEFF (0))$. We denote this step by W ``writing'' in the scheme.\\
\noindent (iii) The quadratures of the atomic memory are now $\pi/2$ rotated; 
$\tilde X^{\text{out}}_{M}  =  \hat P^{\text{out}}_{M}$ and $\tilde P^{\text{out}}_{M}  =  -\hat X^{\text{out}}_{M}$.\\
\noindent (iv) The many-body system is let to evolve freely during an interval of time $t$ such that
$\JEFF(t) = \hat U^\dagger(t) \JEFF \hat U(t)$. Due to the QND character of the Faraday interaction,
$\JEFF(t)$ is unaffected by the first interaction used to map its value to the memory.\\
\noindent (v) At time $t$, a second light beam L$_2$ is sent through both the system and the memory and reads (R)
$\hat X^\text{out}_{L_2}=\hat X^{\text{in}}_{L_2} - \kappa_2 \JEFF(t)+ \kappa_R \tilde P^\text{out}_M$.\\
\noindent (vi) Finally, a balanced homodyne measurement of $\hat X^\text{out'}_{L_2}$ is performed, denoted by D (detection) in Fig.1b. \\

The variance of this observable is the key probe of the protocol, and is given by
\be \label{eq:signal}
[\Delta \hat X^\text{out}_{L_2} ]^2 = \eta(t) + \kappa_T F_{M}(t),
\ee
where we have defined the noise of the signal as $\eta(t)=\mathcal{N}+ \kappa^2_2 [\Delta \JEFF(t) ]^2 + \kappa^2_T/\kappa^2_2 [\Delta \JEFF(0) ]^2$, being $\kappa_T = \kappa_1 \kappa_2 \kappa_R \kappa_W$, and $\mathcal{N}=(1+\kappa^2_R+\kappa^2_R \kappa^2_W)/2$. 
Assuming as usually that the input laser field are in a coherent state and the memory is in coherent spin state, i.e. 
$\aver{\hat X^{\text{in}}}=\aver{\hat P^{\text{in}}}=0$ and $[\Delta \hat X^\text{in}]^2 =[\Delta \hat P^\text{in}]^2=1/2$, the $\mathcal{N}$ noise can be safely neglected, \eg~with $\kappa_1, \kappa_2 \gg \kappa_R, \kappa_W>1$. By performing the two parts of the protocol independently, $\kappa^2_2 [\Delta \JEFF(t) ]^2$ and $\kappa^2_T/\kappa^2_2 [\Delta \JEFF(0) ]^2$ can be measured and thus subtracted. 
After a tedious but elementary derivation, the signal obtained is
\be \label{eq:signalOK}
\begin{split}
F_{M}(t)&= \avg{[ \JEFF(t),\JEFF(0)]_{+} } - 2 \avg{\JEFF(t)}\avg{\JEFF(0)}.
\end{split}
\ee
which provides direct access to the symmetrized two-time dynamical correlation function of $\JEFF$. Using eqn.~\eqref{eq:J} leads to
 $F_{M}(t)= \sum_{n,m} c_n c_m \left[ G_{mn}(t,0) +G_{nm}(0,t)  \right]/N$, where the two point spin time correlation function is
$G_{mn}(t,t') = \langle j_m^z(t) j_n^z(t')\rangle- \langle j_m^z(t)  \rangle \langle j_n^z(t')\rangle$  and the coefficients $c_{n}$
depend explicitly on $\alpha$ and $k$. 

If the state of interest is an eigenstate of the many-body Hamiltonian, the Fourier transform of $F_{M}$ reads $C_{M}(\omega) \propto \sum_n \xi_n [\delta(\omega - (E_n-E_0)) + \delta(\omega + (E_n-E_0))]$ which provides direct information of the energy spectrum. Indeed, $C_{M}(\omega)$ can be related to the symmetric spin dynamical structure factor \cite{VanHove1954}, which is well known to give access to the energy spectrum and the dispersion relation of the system. In this case, $\eta(t)$ is constant and the noise becomes a peak 
at zero frequency and can be thus easily distinguished from the signal with features at finite frequencies.

\begin{figure}
\begin{center}
\includegraphics[width=.8\linewidth]{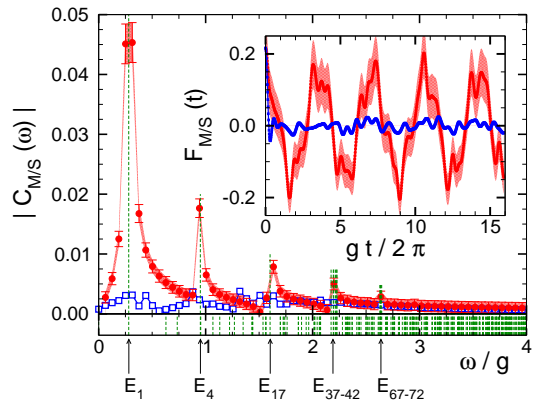}
\caption{(Color online) Fourier transform of $F_{M}$ and $F_{S}$ for Hamiltonian (Eq.~\eqref{eq:Ham}).
The results are obtained with exact diagonalization for a chain of $12$ spins, with a standing wave configuration such that $k=\pi/(2 a)$ and $\alpha=0$. $C_{M}$ (filled red circles and dashed line) and $C_{S}$ (empty blue squares and solid line) are plotted as a function of $\omega$. Vertical dashed green lines show the spectrum of the Hamiltonian. The inset shows the strongly varying $F_{M}$ (red), Eq.~\eqref{eq:signalOK}, and the almost flat $F_{S}$ (blue), Eq.~\eqref{eq:signalNOK},  as a function of time. The uncertainty in the QMAP signal due to experimental limitations is plotted as error bars (losses in the quantum memory lead to a $~5 \%$ reduction of the signal), see the supplementary material for details.   }
\label{Fig:Example}
\end{center}
\end{figure}

To illustrate the differences between $F_{M}$ and $F_{S}$ we study an array of coupled double well superlattices
of $2N$ spins:
\be \label{eq:Ham}
\hat H=  \sum_{n=0}^{N-1}  \left[ g_{1} \SOP_{2n} \cdot \SOP_{2n+1} + g_{2} \SOP_{2n+1} \cdot \SOP_{2n+2}\right],
\ee
where $\SOP=(\hat j^{x}, \hat j^{y}, \hat j^{z})$, $\hat j^{x,y,z}$ are the usual spin-$1/2$ operators, and $g_{1(2)}$ is the coupling between even-odd (odd-even) spins. This Hamiltonian has been implemented with optical superlattices (see~\cite{Barmettler2008,Trotzky2008} and  references therein). In this case $g_{1(2)}=4 t_{1(2)}^{2}/U$, where $t_{1(2)}$ is inter (intra) double well hopping rate, and $U$ the on-site interaction energy, and $g^{-1} \sim 10$ ms which is of the same order of reported storing times ($T$) in quantum memories, as cited before. The condition $gT \gg 1$ is indeed required to resolve the dynamics of the many-body system, which demands either longer storage times or faster dynamics.  Remarkably,  a recent experiment~\cite{Simon2011} simulating antiferromagnetism with tilted optical lattices provides a faster dynamical timescale given by only the tunneling rate, which relaxes the requirements on the memory. We address first the equilibrium case, where $g_{1}=g_{2}=g$ and the initial state is the many-body ground state $\ket{E_{0}}$. In Fig.~\ref{Fig:Example}, the discrete Fourier transform of $F_{S}$ and $F_{M}$ using exact diagonalization are plotted with the parameters given in the caption of the figure (the inset shows $F_{S}$ and $F_{M}$). The signal obtained with the statistical analysis $F_{S}$ is very weak and contains many frequencies, therefore the Fourier transform $C_{S}$ is almost flat. With the use of the quantum memory, much less frequencies are present in the signal and $C_{M}$ shows large peaks at some of the energy levels, including the energy gap to the first excited state.  The robustness of the QMAP method in the presence of experimental limitations and memory loses can be inferred from the error bars in Fig.~\ref{Fig:Example} as explained in the supplementary information. One could also study the non-equilibrium case by considering the system initially prepared in the ground state of the Hamiltonian Eq.~\eqref{eq:Ham} with $g_{1}=g$ and and $g_{2}=0$, that is, a product of singlets between odd-even spins. Then, at $t=0$, the Hamiltonian is quenched to $g_{1}=g_{2}=g$, and the state would evolve accordingly.  Our method 
 could explore thermalization effects characterized by dynamical correlations~\cite{Rossini}, an interesting study which is beyond the scope of this work.

In conclusion, we have shown that by combining a QND light-matter interaction and a quantum memory,
a signal which exploits quantum interference provides direct access to dynamical spin-spin correlations. It is remarkable that in the presence of a many-body Hamiltonian which does not commute with the QND interaction, even the storage of a single quadrature in the quantum memory yields results different from those obtained with a classical memory. The integration of quantum memories in coherent spectroscopy techniques offers unconventional possibilities to manipulate the dynamics of the many body systems by performing conditional feedback operations on the quantum memory ~\cite{Hammerer2010,Mitchell2010}. Particularly appealing are also prospects for detection of multi-time correlations or the study of non-equilibrium physics in strongly correlated systems.

We are grateful to J. I. Cirac, W. D. Phillips, and U. Schneider for useful discussions. We acknowledge support of Alexander von Humboldt Stiftung, EU (AQUTE, NAMEQUAM,MALICIA,QESSENCE), ERC (QUAGATUA), MINCIN (FIS2008-00764,01236), SGR2009-00347, ENB (Project QCCC).

\end{document}